\documentclass{he_symp}
\usepackage{psfig,graphicx,epsfig}
\usepackage{color}
\usepackage{amsmath,amssymb,epic,eepic,array}
\begin{document}
\renewcommand{\FirstPageOfPaper }{ 171}\renewcommand{\LastPageOfPaper }{ 176}


\title{Reviewing Pulsar Spectra}
\author{W. Sieber}
\institute{Hochschule Niederrhein, University of Applied Sciences
             Reinarzstr. 49, D-47805 Krefeld, Germany}
\maketitle
\section{Problems solved?}
Pulsar research must be considered - 35 years after the detection of pulsars - a mature science,
where the basic questions have been raised and discussed. One would hope that many if not all generic and
important problems have found some kind of answer and that scientific work can concentrate now on
specific details requiring more in depth investigation. We know, however, that this picture is not true.
Even well studied areas did in the past not always lead to a general accepted model and some were
investigated at the beginning with enthusiasm but left behind.

     This paper will concentrate on one narrow topic, pulsar radio spectra. It is the attempt to work
out, what features are now generally accepted, but also what features are still in discussion after so many
years of pulsar research. In this sense the paper will enlighten in a personal view those aspects which are
still under discussion.

\section{The spectrum, indicator of the emission mechanism}

The spectrum, i.e. the total emitted energy in specific radio bands, was considered from the
beginning as an easy to measure generic characteristic of the emission mechanism, especially since one
could hope to learn something about an exciting new type of coherent amplification. Comella (1971)
collected and published the first data points from measurements made at Arecibo, a work which was
carried on by Sieber (1973) covering a wider compilation of observations from different observatories.
Later on comprehensive data collections were published by Izvekova et al. (1981) and Lorimer et al.
(1995) who extended the compilation to several hundreds of pulsars, albeit concentrated on few frequency
bands. More recently wide collections of data were published by Malofeev (2000) and Maron et al.
(2000).

     From the beginning it was clear that pulsar spectra are characterizes by a "steep" spectral index,
steeper than e.g. measured for supernova remnants, which explains why they were detected at rather low -
below 1 GHz - frequencies. There were in addition indications that the spectra even steepen in some cases
at higher frequencies beyond 1 GHz, although the measurement uncertainty was high. Undoubtedly, some
pulsars showed a "low-frequency turnover", a decrease of the emitted energy at low frequencies, typically
around 100 MHz, whereas other pulsars had a straight spectrum down to the lowest observable frequencies.
 The cause of this decrease - some kind of absorption in the magnetosphere or a loss of efficiency of
the emission mechanism was discussed - remained unclear.

     The problem with all the measurements was a high variability of the pulsar emission on many
time scales, which resulted in a high uncertainty of the data points and which made it questionable, if a
constant, time invariable spectrum exists at all. Investigations of pulsar spectra have, therefore, to face
this problem and to look into the various possible contributions to the variability.

\section{Variability}
\subsection{Intrinsic variability}

Emission within a pulse during the pulse window is - depending on the pulsar - highly structured
showing sometimes features as short as microseconds (microstructure) or even nanoseconds. Most pulsars
have clearly visible sub-pulses extending over milliseconds, which in some rare cases are well organized
with a shift from pulse to pulse, the so called drifting sub-pulses, Fig. 1.
\begin{figure}
\centerline{\includegraphics[width=7cm]{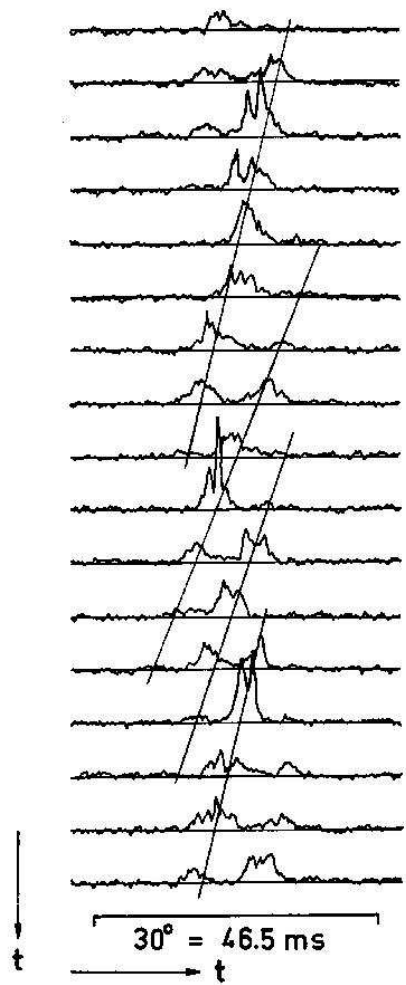}}
\caption{Drifting sub-pulses of PSR B2016+28
\label{image}}
\end{figure}

This effect can now be quantitatively explained by the occurrence of an ExB force in the magnetosphere
working on the out streaming particles. The drift bands cause a well defined periodic modulation (variability)
 of the received emission from pulse to pulse called P3 with characteristic time scale of several
periods. Interestingly, some pulsars exhibit periodic modulations without recognizable drifting sub-pulses,
where the duration extends over many periods, a kind of memory of the radiation mechanism which seems
to be poorly understood so far. 

     Also poorly understood seems to be the producing mechanism for a sudden drop of the emission
to zero (at least below the detection limit) called nulling, an effect which can be found for many pulsars.
These nulls can stretch over many periods, even dominating the emission pattern, so that the emission
occurs only in well defined bursts; an effect, which obviously results in a strong modulation of the
radiation which may extend over many minutes of time.
     
     Measurements show that the mentioned intensity variations are essentially broadband
 ($> 100 \rm MHz$)
 and with high probability related to the emission mechanism or the structure of the magnetosphere
or the surface of the neutron star. The time scales extend from nanoseconds to minutes. It is the hope (we
will come to this point later) that averages of the received signal over many minutes will "smooth out" all
these variations. 

     There is less confidence with respect to another effect, which was called mode switching: In
contrast to the well known behavior that the average of many pulses results in a stable average pulse
profile, some pulsars show sudden variations of the average profile normally correlated with a change in
intensity (Fig. 2). The times of variation are unpredictable and may occur suddenly even after e.g. several
hours (Fig. 3).
\begin{figure}
\centerline{\includegraphics[width=7cm]{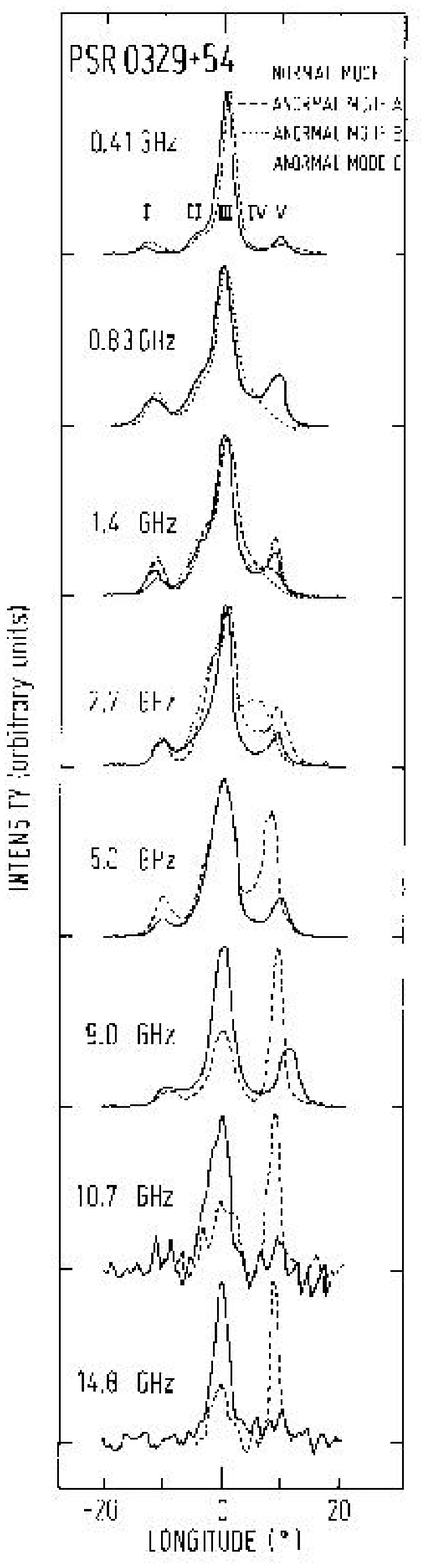}}
\caption{ Mode switching of PSR B0329+54
\label{image}}
\end{figure}
\begin{figure}
\centerline{\includegraphics[width=7cm]{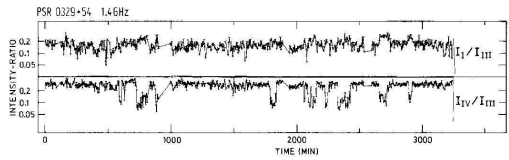}}
\caption{ Mode switching for PSR B0329+54; designation of components explained in Fig. 2
\label{image}}
\end{figure}

For mode switching pulsars, time stable intensity values are obtainable only after some hours of integration, 
depending on the degree of accuracy which one would like to reach (see e.g. the investigation of
Helfand et al. (1975) on the stability of integrated pulse profiles). Nevertheless, the hope is, that the
intrinsic intensity variations can be averaged out by measurements lasting long enough - at least many
minutes, for some pulsars even several hours. Least stable seem to be the mode switching pulsars whose
long-term behavior has not yet been investigated very well.

\subsection{  External influence}

\subsubsection{  Diffractive interstellar scintillation DISS}
     Scheuer (1968) showed as early as 1968 that interstellar scintillation may be the cause of the
modulation of the received pulsar signal on time scales around several minutes, mainly because the small
extent of the emission region makes pulsars ideal "point sources". Rickett (1969) could later proof that
indeed features of strong and weak scintillation can be found in the received emission of pulsars, a
decorrelation of the received signal in frequency and time being the clearest signature of this scintillation
process, now being called diffractive interstellar scintillation DISS. The decorrelation is best represented
by way of "dynamic spectra" (Fig. 4), where the average emission of one or several pulses is analyzed in
a spectrometer and displayed as intensity contours. Successive spectra then depict the development of the
structure in time.
\begin{figure}
\centerline{\includegraphics[width=7cm]{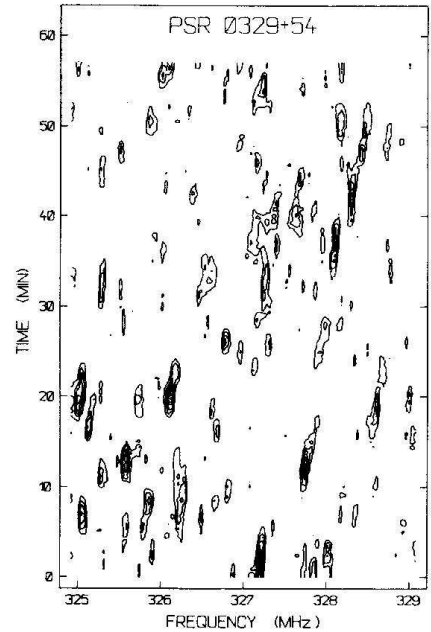}}
\caption{ Dynamic spectrum of PSR B0329+54 at ca. 327 MHz
\label{image}}
\end{figure}

Typical time scales for diffractive scintillation range from seconds to hours and frequency scales from
about 100 kHz to a few 100 MHz. Since scattering parameters are now known for many pulsars, at least
in the strong scintillation regime - albeit by far not for all and not for each frequency range - it is in
principle possible to remove the strong intensity variations caused by DISS. Carefully planned 
observations, extending over much more than the decorrelation bandwidth, and/or with durations much
longer than the decorrelation time should "average out" the variations so that the intrinsic emitted signal
becomes measurable.

     Alternatively one could take many observations of the emitted energy at a certain frequency - even
if bandwidth and duration are smaller than the decorrelation scales - and average these scintillation
modulated signals afterwards. The averaging process should smooth out the variations if enough samples
are taken. Interestingly enough, this was more or less the procedure which was taken intuitively by the
first observers, who just collected the measurements of many authors or observing periods.

\subsubsection{    Refractive interstellar scintillation RISS}
     For years it was known that pulsars show intensity variations even if diffractive scintillation
effects are removed. The cause of the remaining long-term variations was unclear until Sieber (1982)
could show, that they are correlated with the dispersion measure and due to the interstellar medium. The
theoretical foundation was laid by Rickett, Coles and Bourgois in 1984, who showed that focussing and
defocussing by features large compared to the Fresnel scale are the origin, an effect which is now called
refractive interstellar scintillation RISS. The effect is rather broadband and the times scales rather large,
extending perhaps even up to months or years as demonstrated e.g. by long-term observations of 
Stinebring et al. (2000) and Ramesh Bhat et al. (1999). 

     The results of Ramesh Bhat et al. (1999) shook, unfortunately, the confidence that diffractive
scintillation effects may be easily removable by averaging over the decorrelation time or frequency scale,
because they showed that the time and frequency scale is also variable, being influenced by the longer
lasting RISS. Only really long-term observations are apparently able to tell the truth about the emitted
signal strength.

     The measurements of Stinebring et al. (2000) contained also good news: Their observations
demonstrate quite clear that some pulsars, like PSR B0736-40, B1641-45 and B1859+03 exhibit very
stable intensity over many years of observations. This means that the intrinsic emission at least for these
pulsars must be stable. The observations give a clear hint, that the intrinsic emission of perhaps all pulsars
is stable on long terms (intrinsic effects being excepted as demonstrated in Sec. 3.1). So that we have good
hope that we can talk about stable intrinsic (emitted) pulsar radio spectra (on time scales as discussed
above); which means that the original, naive, approach to the problem of pulsar spectra - being based on
the assumption of stable intrinsic spectral behavior - has found a justification. There remains the
problem, that very few pulsars are known so far whose emission has been measured to be stable on long
terms. Far more long-term observations are needed to create a sound basis for accurate pulsar spectra
determinations.

\section{  Spectra}

\subsection{  Spectral index}
     0n the basis of the discussion in the preceeding sections, it seems appropriate to talk about pulsar
spectra as a well-defined intrinsic characteristic of pulsars. Observations of the first years showed already
the steep character of the spectra (Sieber 1973), with an exponent of -1.62 for the high-frequency, power
law part of the spectrum (not taking into account the low-frequency turnover found for some pulsars and
without considering the high-frequency steepening). These first investigations were based on few sources
(27), a number, which was increased later on by Izvekova et al. (1981) (52 sources) and then in a big step
to 280 by Lorimer et al. (1995), Malofeev (1996, 2000) and Maron et el. (2000).

\begin{table}
      \caption{Compilations of pulsar spetra}
         \label{KapSou}
      \[
         \begin{tabular*}{\linewidth}{p{0.25\linewidth} p{0.125\linewidth} p{0.125\linewidth} p{0.3\linewidth}}
            \noalign{\smallskip}
             \hline
         \noalign{\smallskip}
Author &Number of  Pulsars  &Mean  Spectral Index &  Remarks\\
            \noalign{\smallskip}
             \hline                         
Sieber 1973      &        27  &           - 1.62 & \\
Izvekova et al.  1981  &        52  &           - 2.0 & \\
Lorimer et al. 1995 &     280 &            - 1.6 &         small frequency range\\
Malofeev 2000&           340  &          - 1.71&         wide frequency range\\
Maron et al. 2000&              &         - 1.8   &       more high frequency values\\
            \noalign{\smallskip}
             \hline
        \end{tabular*}
      \]
\end{table}

     The computed mean spectral indices vary for these investigations between -1.6 and -2.0, underlining 
the steep character of the emitted spectra, but showing also a remaining uncertainty in the determination 
of the average value (which makes it not very sensible to give error bars). There is a wide
scatter of measured spectral indices apparent for individual pulsars, ranging from 0.0 (or even slightly
positive) to -3.1, which may explain also  the scatter in the compiled average spectral index. There is also
some uncertainty due to the fact that the frequency range used for the determination of the exponent is
often limited either by lack of data or because one has to exclude either the low-frequency or the high--frequency 
turnover.

     The spectral index is only weakly correlated with any other measurable pulsar parameter, as e.g.
P, dP/dt etc., the clearest correlation being given to the characteristic age (Lorimer et al. 1995). It remains
to theories of the emission mechanism to explain the physical reasons for the 
differences in spectral index   for different
pulsars (See however section 6).

\subsection{  Low-frequency turnover}
     A high percentage of those (normal) pulsars, where flux values below 100 MHz have been
measured, show a low-frequency drop (turnover) in intensity (see Fig. 5); and it is known that correlations
between for example the maximum frequency and the pulsar period exist. 

\begin{figure}
\centerline{\includegraphics[width=7cm]{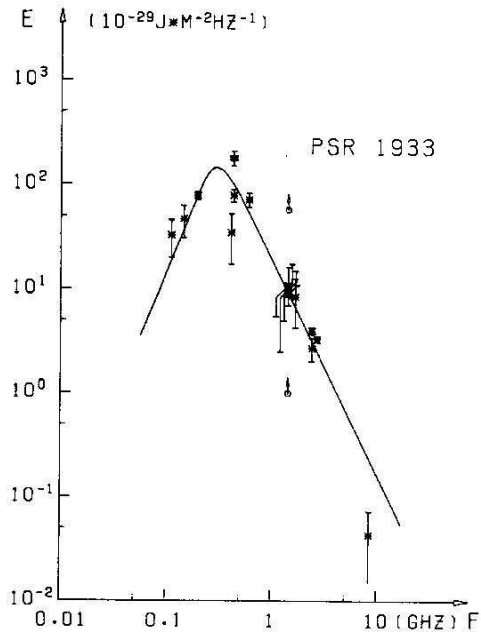}}
\caption{ Low-frequency turnover of PSR B1933+16
\label{image}}
\end{figure}
     Several mechanism have been proposed to explain this drop in intensity, often related to the
radiation mechanism. Since the debate on the most probable emission mechanism is still ongoing, there is
also still no definite answer to the cause of this low-frequency turnover.

     It is interesting to note, that some pulsars on the contrary show definitely no low-frequency
turnover, although accurate measurements are available for them down to very low frequencies (10 MHz).
Famous examples are the Crab pulsar (Fig. 6) and the millisecond pulsar PSR B1937+214 (Erickson \&
Mohany 1985).
\begin{figure}
\centerline{\includegraphics[width=7cm]{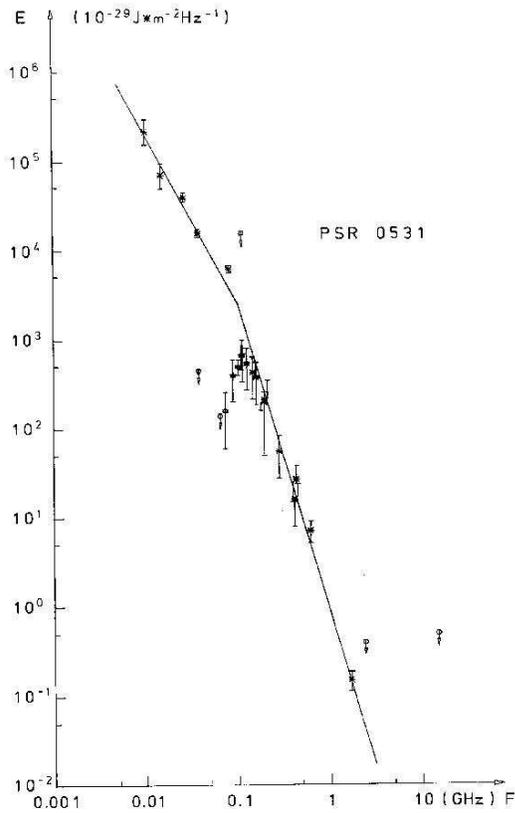}}
\caption{Spectrum of PSR B0531+21, the Crab pulsar
\label{image}}
\end{figure}
     Explanations for the low-frequency turnover should also clearly state, why some pulsars obviously 
show increasing intensity down to very low frequencies.

\subsection{  Break in the power law spectrum}
     There were already at the beginning of the investigation of pulsar spectra indications, that also the
high-frequency part of the spectrum may not completely be describable by just one power law. Very few
pulsars showed indications of a steepening of the spectrum (see e.g. Fig. 7), which could be fitted by two
power laws. The fact that pulsar emission is highly variable on many time scales (Sec. 3), makes it
difficult to firmly proof such a break in the spectrum and to separate clearly intrinsic effects from e.g.
long-time refraction effects. It should, however, be mentioned, that a break in the spectrum is a natural
consequence of some theoretical radiation emission mechanisms; see e.g. Ochelkov \& Usov (1984).
\begin{figure}
\centerline{\includegraphics[width=7cm]{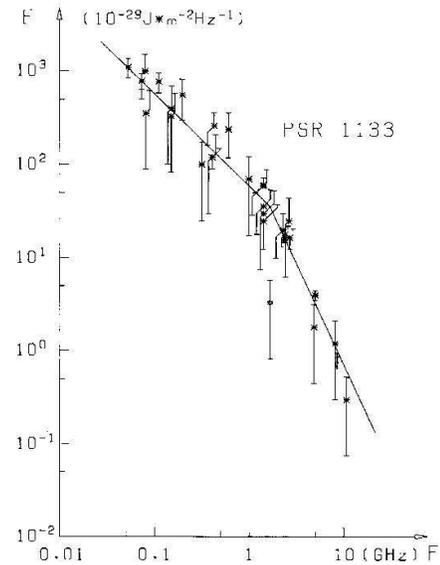}}
\caption{Spectrum of PSR B1133+16
\label{image}}
\end{figure}

     Most interesting appears also the fact that some pulsars show obviously an increase of their
emission intensity at very high frequencies; this effect will be discussed in detail in the contribution of
Wielebinski (2002) in these proceedings.

\section{  Millisecond pulsars}
     The detection of millisecond pulsars opened up a new era in the investigation of pulsars. Millisecond
 pulsars occupy a different region in the parameter space and build therefore a severe test for every
theory of radio emission. Comparisons between the characteristics of normal pulsars and millisecond
pulsars were therefore eagerly expected. Kramer et al. (1998) published the spectra of 32 millisecond
pulsars and found an exponent of   = -1.8 if no distance limit was applied. Nearer pulsars (distance less
then 1.5 kpc) showed a bit flatter spectrum with   = - 1.6. They compared this result with 346 normal
pulsars, which showed an exponent of   = -1.6 for all distances and of   = - 1.7 for distances $< 1.5\rm\ kpc$.
In correspondence to the result of Toscano et al. (1998), who measured an index of   = -1.9 for southern
millisecond pulsars, one comes to the conclusion that there exists no real difference in the spectral index
between normal pulsars and millisecond pulsars. It seems that the same radiation mechanism works in
both environments.

\subsection{  Low-frequency turnover}
     Kuzmin \& Losovsky (2001) were able to extend the spectral frequency range of quite a collection
of millisecond pulsars down to 100 MHz. Interesting enough no low-frequency turnover (drop at low
frequencies) was detected, with one exception, PSR J1012+5307. Together with the fact that the millisecond 
pulsar PSR B1937+214 shows a straight spectrum down to 10 MHz, there is the indication, that
low-frequency turnovers might be rarer in millisecond pulsars than in normal pulsars.

\section{  Geometry}
     The geometry of the hollow-cone must have a direct impact on the spectrum, if the radio emission
is - as we believe - concentrated in a hollow cone with a frequency dependent width. Kramer et al. (1994)
and Sieber (1997) could demonstrate that the observed spectrum may be indeed steeper than the emitted
one, the steepening being caused by the shrinking of the hollow-cone width at high frequencies.

     An emitted spectrum with exponent   = -1.5 could - depending on the specific geometry - be
observed as a spectrum with   = -2. This would mean, that the radiation mechanism itself must not
produce such steep spectra as we observe them. 

\begin{figure}
\centerline{\includegraphics[width=7cm]{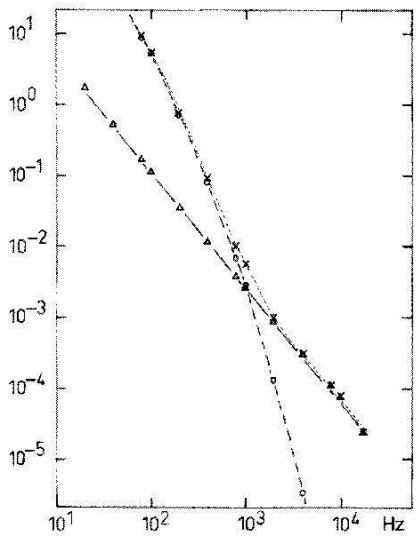}}
\caption{Simulation of the steepening of the spectrum (Sieber 1997): A gaussian core component is assumed
(circles) and a cone component (triangles). The measured intensity of the core component and of the whole profile
(crosses) drops rapidly just due to the shrinking of the profile width at high frequencies
\label{image}}
\end{figure}

     In addition, the received intensity may depend on several more geometrical factors, e.g. the
curvature radius of the field lines, the emission hight over the neutron star surface and the orientation of
magnetic field axis and rotation axis. Kuzmin \& Solov'ev (1986) worked out an curvature radiation
emission model which reflects quite well as an example the spectral characteristics of PSR B1133+16,
including the low-frequency turnover and taking into account all these geometrical effects.

     It appears obvious that geometrical effects influence the shape of the received spectrum; but it is
much harder to determine the exact impact in individual cases, since then the precise geometry has to be
known, e.g. the angle between rotation axis and magnetic field  , as well as the impact angle  . Normally
the geometry is derived from measurements of the polarization angle sweep across the profile; these
measurements are, however, often spoiled by orthogonal polarization angle jumps and may give unreliable
results. Most promising are polarization angle histograms which often allow to determine at least the angle
sweep with good precision.

\section{  Conclusions}
     From the arguments presented one is led to the conclusion that astonishingly few problems found
an answer so far. One of these is obviously the question of the stability of the intrinsic spectrum which
seems to be settled in the sense, that we can now indeed assume a stable emitted spectrum being heavily
modulated by transmission effects on all time scales extending up to months. Also the intrinsic 
variations manifested as drifting sub-pulses and P3 modulation can now be explained more and more in the
framework of geometrical mapping to the polar cap of the neutron star.

     To the still debated problems belongs e.g. the question of the cause of the low-frequency turnover
in the spectra of pulsars - normal and millisecond - the cause of the high-frequency break in some spectra
and the turn-up at extreme high frequencies. And finally it would be of great help if the geometry of the
emission geometry of more and more pulsars could be determined with higher confidence.

Acknowledgments: The support of the Wilhelm und Else Heraeus-Stiftung is very much acknowledged.
This work was also supported by INTAS grant no. 2000-849.


\clearpage


\begin{thebibliography}{}

\bibitem{1}{Comella J.M., 1971, Ph.D. thesis}

\bibitem{2}{Erickson W.C., Mahoney M.J., 1985, ApJ 299, L29}

\bibitem{3}{Helfand D.J., Manchester R.N., Taylor J.H.: observations of pulsar radio emission. III. Stability of
integrated profiles, 1975 ApJ 198, 661-670}

\bibitem{4}{Izvekova V.A., Kuzmin A.D., Malofeev V.M. et al., 1981, Ap\&SS 78, 45}

\bibitem{5}{Kramer M., Wielebinski R, Jessner A., Gil J.A., Seiradakis J.H.: 1994, A\&ASS 107, 515}

\bibitem{6}{Kramer M., Xilouris K.M., Lorimer D.R. Doroshenko O., Jessner A., Wielebinski R., Wolszczan A.,
Camilo F.: The characteristics of millisecond pulsar emission. I. Spectra, pulse shapes, and the beaming
fraction, 1998 ApJ 501, 270-285
}
\bibitem{7}{Kuzmin A.D., Losovsky B.Ya.: No Low Frequency Turn-over in the Spectra of Millisecond Pulsars, 2001
A\&A} 

\bibitem{8}{Kuzmin A.D., Solovev A.G.: Model pulsar radio spectra and pulse profiles, 1986, Sov. Astron. 30, 38
}
\bibitem{9}{Lorimer D.R., Yates, J.A., Lyne A.G., Gould D.M.: Multifrequency flux density measurements of 280
pulsars, 1995 MNRAS 273, 411-421

\bibitem{10}{Malofeev V.M., 1996, PASPC 105, 271}

\bibitem{11}{Malofeev V.M., 2000, PASPC 202, 221 }

\bibitem{12}{Maron O., Kijak J., Kramer M., Wielebinski R.: Pulsar spectra of radio emission, 2000, A\&ASS 147,
195-203}

\bibitem{13}{Ochelkov Yu.P., Usov V.V.: The nature of low-frequency cutoffs in the radio emission spectra of pulsars,
1984, Nature 309, 332-333}

\bibitem{14}{Ramesh Bhat N.D., Pramesh Rao A., Yashwant Gupta: Long-term scintillation studies of pulsars. I.
Observations and basic results, 1999, ApJS 121, 483-513}

\bibitem{15}{Rickett B.J., 1969 Nature 221, 158}

\bibitem{16}{Rickett B.J., Coles W.A., Bourgois G., 1984, A\&A 134, 390}

\bibitem{17}{Scheuer P.A.G., 1968 Nature 218, 920}

\bibitem{18}{Sieber W.: Pulsar spectra: A summary, 1973 A\&A 28, 237-252}

\bibitem{19}{Sieber W.: Causal relationship between pulsar long-term intensity variations and the interstellar medium,
1982, A\&A 113, 311}

\bibitem{20}{Sieber W.: Geometrical effects on radio pulsar profiles and spectra, 1997, A\&A 321, 519-522
}
\bibitem{21}{Stinebring D.R., Smirnova T.V., Hankins T.H., Hovis J.S., Kaspi V.M., Kempner J.C., Myers E., Nice
D.J.: Five years of pulsar flux density monitoring: refractive scintillation and the interstellar medium,
2000, ApJ 539, 300-316}

\bibitem{22}{Toscano M., Bailes M., Manchester R.N. et al., 1998, ApJ 506, 863}

\bibitem{23}{Wielebinski R.: Characteristics of (normal) radio pulsars, in: Becker et al. (eds.), Neutron Stars, Pulsars
and Supernova Remnants, Proceedings of the 270 WE-Heraeus-Seminar 2002}
}
\end{thebibliography}
\end{document}